ORIGINAL RESEARCH

# Social Co-OS: Cyber-human social Co-operating system

Takeshi Kato[1] | Yasuyuki Kudo[2] | Junichi Miyakoshi[2] | Misa Owa[2] | Yasuhiro Asa[2] | Takashi Numata[2] | Ryuji Mine[2] | Hiroyuki Mizuno[2]

[1]Hitachi Kyoto University Laboratory, Open Innovation Institute, Kyoto University, Kyoto, Japan

[2]Hitachi Kyoto University Laboratory, Center for Exploratory Research, Hitachi Ltd, Kyoto, Japan

**Correspondence**
Takeshi Kato, Hitachi Kyoto University Laboratory, Open Innovation Institute, Kyoto University, Yoshidahonmachi, Sakyo-ku, Kyoto-shi, Kyoto, 606-8501, Japan.
Email: kato.takeshi.3u@kyoto-u.ac.jp

**Abstract**

The novel concept of a Cyber-Human Social System (CHSS) and a diverse and pluralistic 'mixed-life society' is proposed, wherein cyber and human societies commit to each other. This concept enhances the Cyber-Physical System (CPS), which is associated with the current Society 5.0, a social vision realised through the fusion of cyber (virtual) and physical (real) spaces following information society (Society 4.0 and Industry 4.0). Moreover, the CHSS enhances the Human-CPS, the Human-in-the-Loop CPS (HiLCPS), and the Cyber-Human System by intervening in individual behaviour pro-socially and supporting consensus building. As a form of architecture that embodies the CHSS concept, the Cyber-Human Social Co-Operating System (Social Co-OS) that combines cyber and human societies is shown. In this architecture, the cyber and human systems cooperate through the fast loop (operation and administration) and slow loop (consensus and politics). Furthermore, the technical content and current implementation of the basic functions of the Social Co-OS are described. These functions consist of individual behavioural diagnostics, interventions in the fast loop, group decision diagnostics and consensus building in the slow loop. Subsequently, this system will contribute to mutual aid communities and platform cooperatives.

## 1 | INTRODUCTION

Society 5.0 is the future society advocated in the Japanese fifth Science and Technology Basic Plan following the hunting (Society 1.0), agricultural (Society 2.0), industrial (Society 3.0), and information societies (Society 4.0 and Industry 4.0). It is a human-centred society that balances economic advancement with the resolution of social problems through a system that integrates cyber space (virtual space) and physical space (real space) [1]. To resolve the disparities and inequalities cited by Society 5.0, the United Nations' Sustainable Developmental Goals (SDGs) [2] and social challenges such as environmental and energy-related issues, there is a need to revise the definition of the ideal human society and its relationship with the cyber (information systems) [3].

Society 5.0, as we envision it today, involves Cyber-Physical Systems (CPS) that produce a fusion of big data and artificial intelligence (AI) in the cyberspace with the Internet of Things (IoT) and robots in physical space, which attempts to reproduce the digital twins of real humans and society in the cyberspace. However, considering the theory-ladenness of observation, it is not possible to convert everything into data. Moreover, humans do not necessarily act rationally in accordance with the data, as described by the notion of bounded rationality. Since it is infeasible to reproduce a complete digital twin, our conception of Society 5.0, assuming the optimisation of the entire social system with the use of CPS [1], must be revised.

Studies related to human-incorporated CPS tend to position the Cyber-Human System (CHS) as complementary to CPS [4]. Additionally, the Human-Cyber-Physical System (HCPS) and Human-In-The-Loop (HITL) are positioned in the gradual combination of CPS and CHS.

According to Liu and Wang's survey paper on HCPS [5], it can be divided into three major types. The first type treats human elements as physical entities and controls them using CPS [6]. The second type integrates the model of human psychology and behaviour with the CPS model to improve the performance of the manufacturing and work systems using the human labour force and work efficiency [7, 8]. The third type integrates social networks and CPS, and leverages social information to improve healthcare applications and transportation systems [9, 10]. Additionally, according to some representative papers on Human-in-the-Loop CPS (HiLCPS) [11–13], the HiLCPS incorporates human sensing information and models into CPS and improves the performance of the automatic and autonomous system.

The Cyber-Human System, which focusses more on humans than CPS, is positioned above the three-dimensional space of humans, computers, and the environment [14]. For example, according to several papers on CHS [15–17], it improves security through individual decision-making models, monitors individual mental health with smart devices, and improves performance in workers' remote collaboration.

To resolve the problems of human society targeted by Society 5.0 and SDGs, the CPS needs to be involved in the behaviour and decision-making of individuals and groups pro-socially. Furthermore, it should intervene and support the human society itself. The traditional HCPS and HiLCPS use humans to improve the performance of physical systems, with emphasis on the physical system side, while the traditional CHS focusses on improving personal performance. In contrast, the revised Society 5.0 uses cyber systems to improve human society, with emphasis on the human society side. However, as mentioned above, it is impossible to optimise human society in one go by the cyber system; thus, it is necessary to envision a system in which the cyber system gradually improves while interacting with the human society.

We aim to provide a new cooperating system in which cyber and human societies are connected, replacing the prevailing notion of Society 5.0. In this context, rather than emphasising on economic values behind social issues, pluralistic values, including social and environmental values, diversity of individual human values, and social norms and ethics that influence these values need to be considered. With these considerations, we will aim for a 'mixed-life society' in which people live by entrusting each other through the cyber system [18].

In this paper, we propose the schema of a system that connects cyber and human societies to enhance HCPS, HiLCPS, and CHS in Section 2, present a cooperating system between the two as architecture in Section 3, show the content of basic functions of cooperating systems and current implementation outcomes in Section 4, discuss the ethical challenges of basic functions in Section 5, and describe the conclusions and future perspectives in Section 6.

## 2 | CONCEPT: CYBER-HUMAN SOCIAL SYSTEM

In Society 5.0, IoT connects various objects and people and processes the big data obtained using information technology and AI. This enables real-time analysis and dynamic interventions for human society.

Traditionally, the paradigm of social knowledge has relied on non-real-time physics models (data analysis → explanatory → prediction) and historical models (history analysis → futuristic insight → prevention). However, in Society 5.0, the focus has shifted to real-time clinician models (diagnostic → prognostic and predictive → interventions) [19]. For this reason, social interventions have traditionally been based on posteriori values and judgements. Nonetheless, in the future, it will become necessary to pre-incorporate the process of value judgement and decision-making in a priori to build a cyclical and dynamic relationship between cyber and human societies.

Humans have an 'extended mind' beyond the brain and in the 'external scaffolds' of the body and environment [20]. They are not solely independent existences consisting of the self as 'I'. Rather, as the self as 'We', they include multiple agents such as the body, others, tools, and the environment [18]. From this perspective, cyber is also one of the 'external scaffolds' and one agent that 'We' is composed of.

Thus, we have proposed the Cyber-Human Social System (CHSS) as an extended society, in which cyber and human societies are nested with each other, where the 'We' model operates in cyberspace while the cyber is incorporated into a part of the human society [21, 22]. To emphasise the social human as 'We' rather than the individual human as 'I', we decided to rename it CHSS instead of CHS. As previously mentioned, the model is an incomplete digital twin. Thus, it needs to be updated periodically in a cyclical relationship with human society.

In other words, as shown in Figure 1b, a 'We' model is created from the human society, upon which interventions are delivered from the cyber to the human society. Consequently, the outcomes of the interventions are recursively reflected in the 'We' model and subsequent new interventions. Interventions are based on social norms through social consensus, but their social norms are reviewed through a process of consensus building and group decision making in response to changes in relationships and the environment. Thus, in CHSS, cyber and human societies are committed to each other.

The Cyber-Human Social System is similar to the second type of HCPS and the CHS since it uses a human model. It is similar to the third type of HCPS and the HiLCPS, as it uses social and human sensing information to generate human models. However, CHSS differs from the traditional HCPS, HiLCPS, and CHS, as it dynamically circulates interventions based on social norms and consensus building to review those norms using a human social model called 'We.' As shown in Figure 1b, CHSS includes CPS and CHS and further co-operates between cyber and human society.

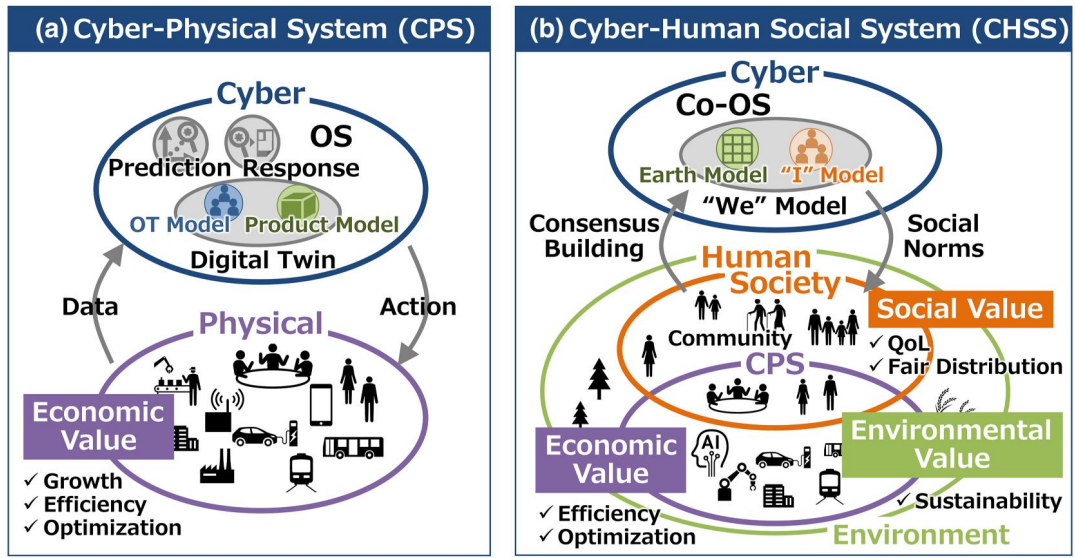

**FIGURE 1** (a) Cyber-physical and (b) cyber-human social systems

## 3 | ARCHITECTURE: CYBER-HUMAN SOCIAL CO-OPERATING SYSTEM (SOCIAL CO-OS)

According to Luhmann's social systems theory [23], the social system is an autopoietic (self-organising) system composed of a circular network of communication, which creates a social order on its own through communication (information, utterance, and understanding).

According to the social institutions theory of Aoki [24] or Pillath and Boldyrev [25], social systems are cyclical systems in which an individual's behavioural orientation triggers strategic behaviours, collectively generating a recursive state of mutual behaviours. This, in turn, generates a social institution (symbolic system) that feeds back into the cycle by triggering an individual's behavioural orientation.

Together with the social systems theory and social institutions theory, the social system is considered a dynamic circulatory system consisting of three hierarchies of individual behaviour, inter-individual interaction, and institutional formation [26]. In other words, inter-individual interactions (communication) of individual behaviours may create social order, which is further systematised into symbols to form the institution.

According to Brandom and Heath's theories [27, 28], social norms arise from mutual expectations and the approval of inter-individual interactions, which form the basis of social order and systems. Resolving social issues requires the cultivation of social norms and ethics. Within this process, the cyber should play a role in CHSS to provide normative interventions and support based on data collected at the three hierarchies of the social system.

According to Kahneman's dual-process theory [29], human information processing consists of an automatic and intuitive fast system and a deliberative and inferential slow system. According to Wilson [30], the social system consists of a fast operation in administration and a slow discussion in politics, with interventions being conducted to maintain social order in operation and consensus building being conducted to form a social institution in the discussion. Considering these, the role played by the cyber is to intervene normatively in the inter-individual interactions in fast operation and support institutional formation in slow discussions.

The concepts explained above constitute the fundamental background of the architecture of the Cyber-Human Social Co-Operating System (Social Co-OS), which embodies the CHSS-concepts, as shown in Figure 2. The Social Co-OS consists of a fast loop responsible for administration and operation and a slow loop responsible for politics and consensus building. This is also regarded as a second-order autopoietic system. In the fast loop, the cyber collects and diagnoses individual behavioural data and intervenes in individual behaviours in the place of inter-individual interactions to generate social norms. In the slow loop, the cyber diagnoses group intention data related to prognostic predictions of pluralistic values and supports consensus building regarding social choices. The social order in the fast loop is maintained based on consensus in the slow loop, and this consensus is updated accordingly in response to changes in the order in the fast loop. Here, 'Co-Operating' carries dual meanings of cooperation between cyber and human societies and that between the fast and slow loops.

The Social Co-OS can be said to be a type of HiLCPS as it interposes humans in a loop; however, there are differences between the two. As mentioned in Section 1, in the traditional HiLCPS, the performance of the physical system is improved by the intervention of the cyber system and through incorporating the human model to the physical system. Conversely, in Social Co-OS, the problems of human society will be resolved by the intervention of the cyber system to the human social system. Furthermore, the Social Co-OS has a double loop of fast and slow, reflecting the construction of human thinking and human society, which differs from the traditional HiLCPS.

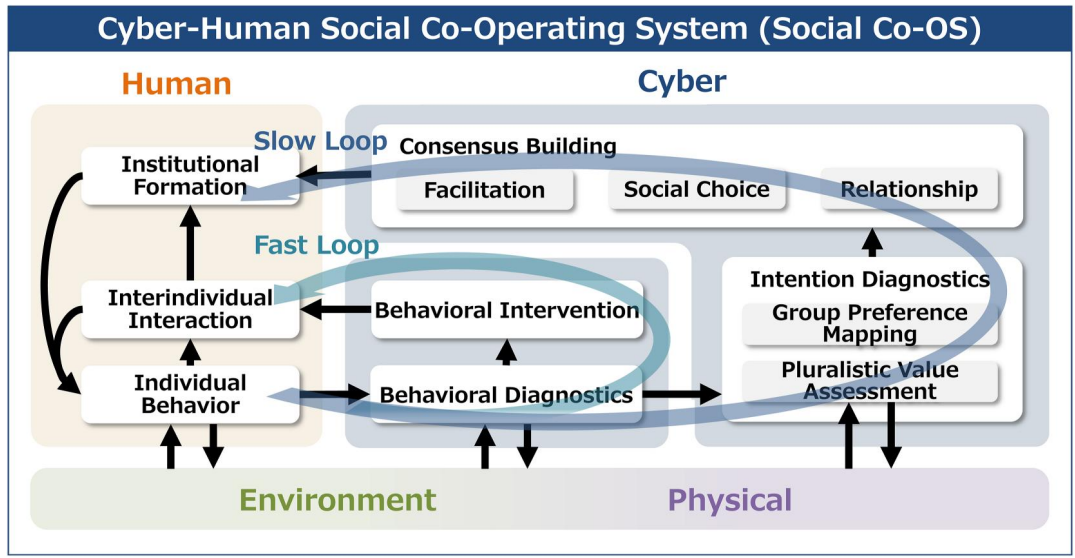

**FIGURE 2** Cyber-human social co-operating system (social Co-OS)

## 4 | BASIC FUNCTIONS OF SOCIAL CO-OS

The basic functions of Social Co-OS consist of the fast loop: individual behavioural diagnostics, individual behavioural interventions, and the slow loop: group intention diagnostics and group consensus building, as illustrated in Figure 2. In this section, we describe the basic functions of these technologies.

### 4.1 | Fast loop: Individual behavioural diagnostics

Human will and behaviour are influenced by the economic utility and social norms [28]. Individual decision-making is interpreted as a balance between utility and norm. In mathematical models, it is interpreted as a choice in the cross-point between utility and norm functions, allowing both functions to be obtained from behavioural data [31]. This approach explains, for example, the convenience of power consumption and the normativeness of energy savings, Gross Domestic Product, and $CO_2$ emissions, and the relationship between resource productivity (sustainability) and fairness (Gini coefficient) [32].

In addition to the binomial relationship between utility and norms, individual behaviour is also linked to personality (psychological traits). Moreover, alongside common questionnaire surveys, the methods for examining personality include estimation from behavioural data. Therefore, we attempted to estimate personality from behaviours in daily living in individual behavioural diagnostics of Social Co-OS [33]. Specifically, power sensors and motion sensors were attached to the simulated living environment shown in Figure 3, and the correlation between the behavioural data and personality was analysed.

Power sensors were placed in seven locations (blue areas in Figure 3) to measure the amount of power used or the time of use during every 60 min, in which the subjects stayed in various areas of the house, as shown in Figure 3. Motion sensors were installed at five locations (orange points in Figure 3), and the maximum and total amount of movement, rotation, and direction change was calculated from three time-series data of acceleration, angular velocity, and geomagnetism for 60 min. Additionally, a principal component analysis was performed for seven use indices (seven sites) and 30 motion indices (five sites × three types × maximum/total), and the scores from the first to the fifth principal components were used as composite indices. As mentioned above, indices of 37 individual indices (7 types + 30 types) and 10 composite indices (use/movement × 5 components), or a total of 47 types, were calculated as life behaviour indicators.

A digitised questionnaire on the Big Five personality traits [Neuroticism-Extraversion-Openness Five-Factor Inventory on Neuroticism, Extraversion, Openness, Agreeableness, and Conscientiousness; NEO Five Factor Inventory (NEO-FFI)] was used to assess personality. Subjects were asked to answer 60 questions on a 5-point Likert scale. Their responses were converted to standard scores using the percentile conversion table of the Japanese version of the NEO-FFI Manual [34], which served as an index of personality. Pearson's product-moment correlation coefficient was used for correlation analysis between these five indices and 47 life behaviour indices. Significance was assessed at a significance level of $p < 0.05$.

The details of the experimental results are given in the literature [33], but the most strongly correlated life behaviour indices for each of the five personality indices are shown in Table 1. Regarding all traits excluding conscientiousness—neuroticism, extraversion, openness, and agreeableness—personality estimation from life behaviours proves to be useful. Correlation graphs for neuroticism, extraversion, openness,

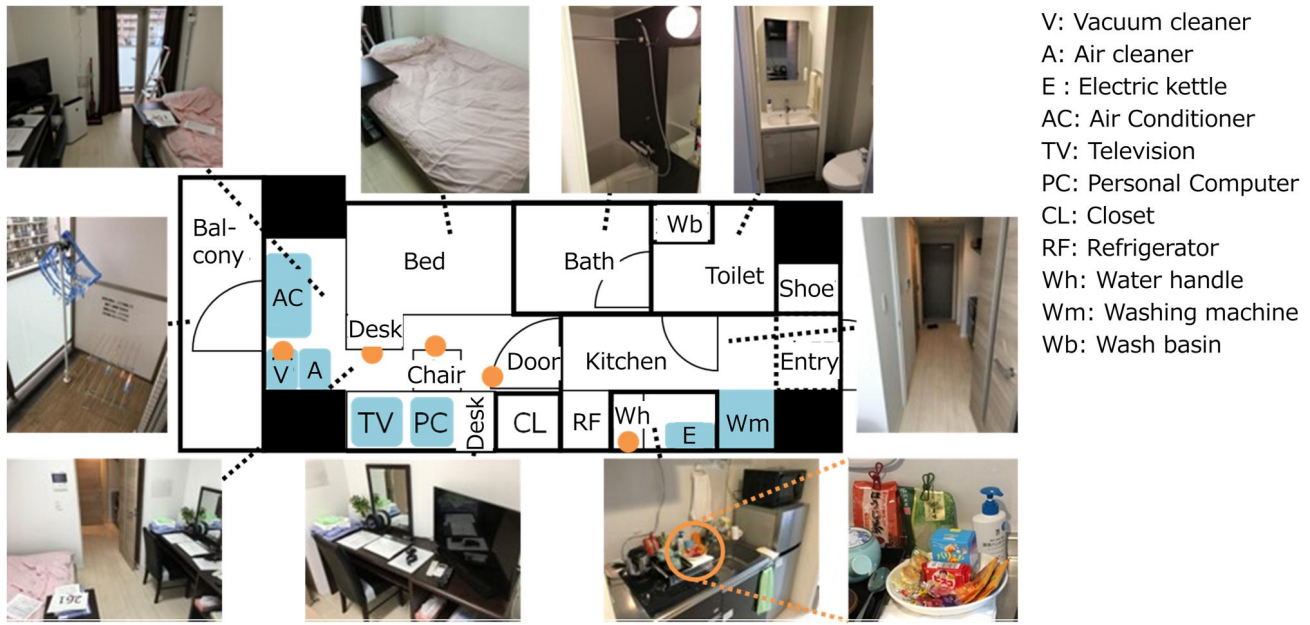

**FIGURE 3** Environment for measuring living behaviour

**TABLE 1** Personality and living behaviour index

| Personality | Living behaviour index | Correlation coefficient |
| --- | --- | --- |
| Neuroticism | Total movement of the small desk | 0.49* |
| Extraversion | Maximum change of the chair | −0.28* |
| Openness | Third principal component of home appliance usage | 0.37* |
| Agreeableness | Maximum rotation of the water handle | −0.32* |
| Conscientiousness | Maximum rotation of the small desk | −0.21 |

*$p < 0.05$.

and agreeableness, which showed significant correlations, are shown in Figure 4. Neuroticism is associated with measures such as effortful control of behavioural control, while agreeableness is associated with social value orientation on prosociality. Thus, we believe that lifestyle behaviour-based individual behavioural diagnostics can be used for normative interventions in fast loops.

## 4.2 | Fast loop: Individual behavioural intervention

Individual behavioural interventions generally include institutional and psychological approaches, but the latter is adopted in the fast loop as the former corresponds to the slow loop of Social Co-OS. Widely recognised traditional psychological approaches include nudging, value-sensitive design, and positive computing. However, we decided to investigate interventions based on a meta-analysis of various existing social psychology experimental data [35].

Specifically, the social dilemmas were taken as sample problems, and data from 700 related psychological experiments were encoded. Machine learning was conducted using it as training data, and the ranking and effects of the intervention measures promoting cooperative behaviours were derived. In machine learning, a new knowledge-embedded neural network (KeNN), as shown in Figure 5, is embedded including six psychological determinant models for social dilemmas. Furthermore, 33 encoding feature vectors were inputs among individual traits, such as personality and gender, and they were connected to KeNN as interactions.

The literature [35] provides detailed descriptions of the KeNN. However, the predictive performance of the KeNN and the results of the determinant score analyses are shown in Figure 6. The horizontal axis in (a) is the cooperative behaviour rate observed in the psychological experiment, and the vertical axis represents the cooperative behaviour rate predicted by KeNN. The correlation coefficient $R$ between the two is 0.79, indicating that the KeNN has adequate predictive performance. Figure 6b illustrates the determinant scores obtained by KeNN for some social dilemma issues in the radar charts.

Normal neural networks are fully connected. Consequently, these networks cannot analyse the determinants of cooperative behaviour, whereas KeNNs can do so. Additionally, by analysing the main characteristics associated with the determinants, the ranking and effects of the interventions can be derived, as shown in Figure 7.

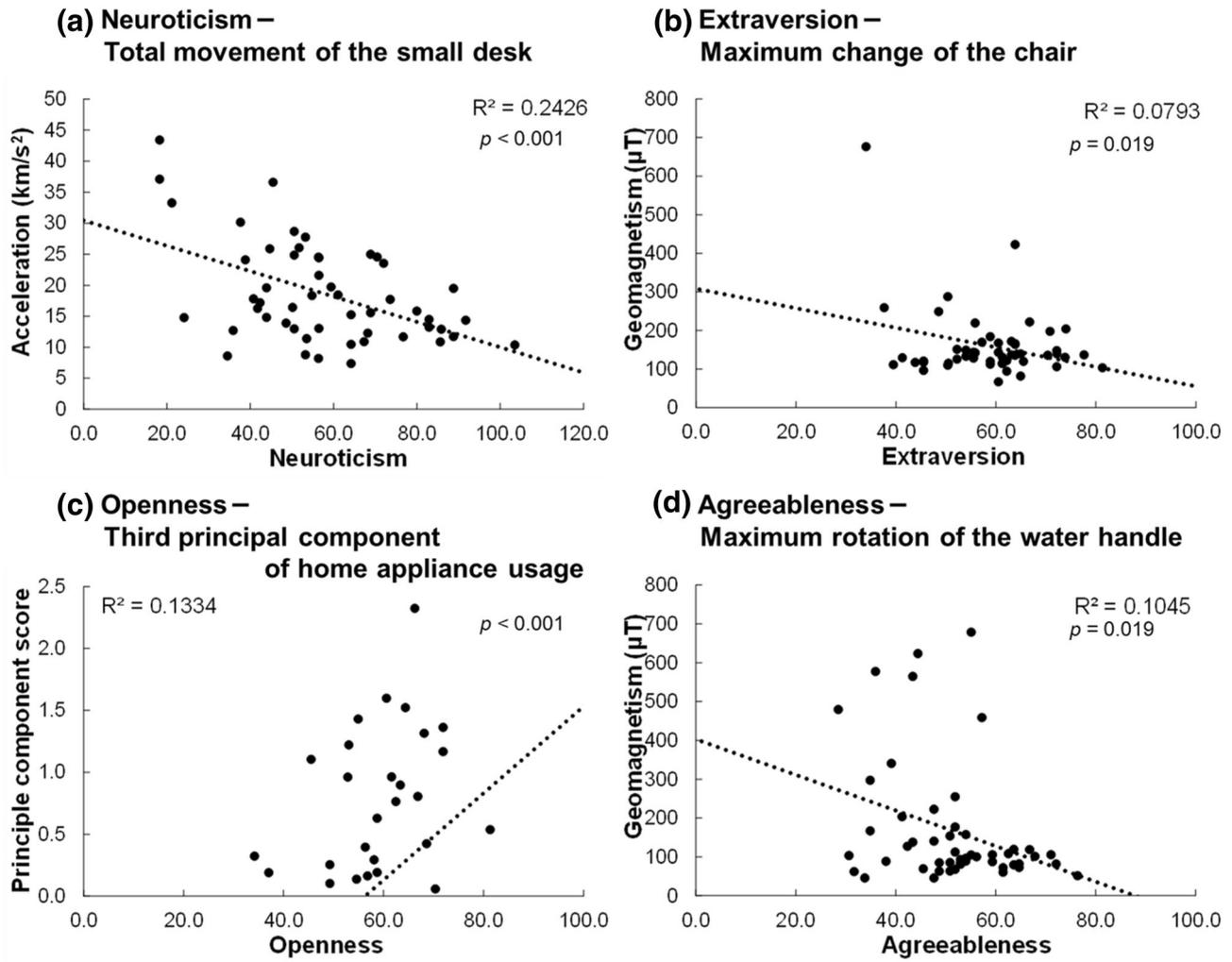

**FIGURE 4** Correlation graph of personality and living behaviour

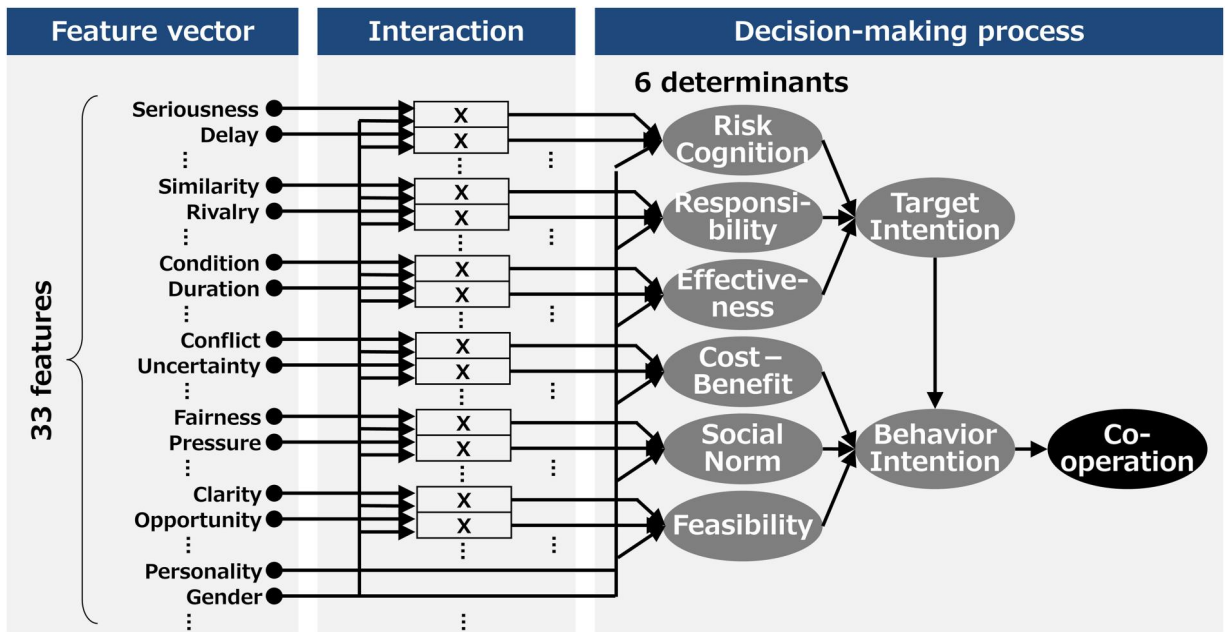

**FIGURE 5** Knowledge-embedded neural network (KeNN)

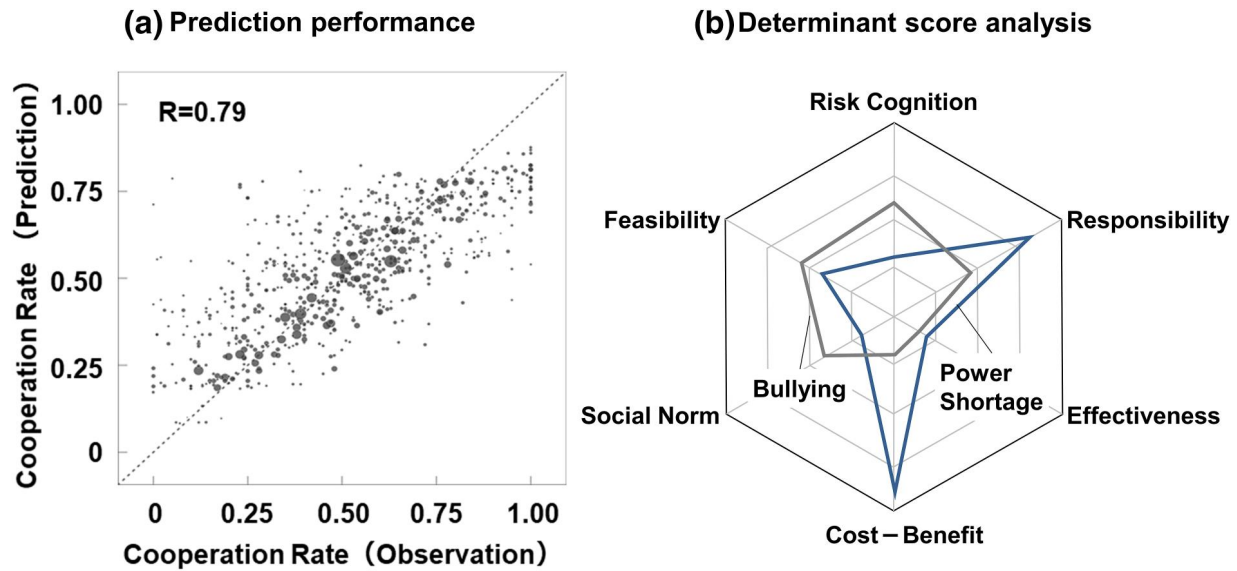

**FIGURE 6** Prediction performance of cooperative rate and determinant score analysis

In the fast loop, we believe that connecting the individual behavioural diagnostics personality information presented in the previous section to the individual behavioural interventions presented in this section (i.e. by inputting KeNN interactions) will allow for more effective interventions, which are customised to the individual context, as illustrated in Figure 8. Behavioural predictions and interventions with KeNN correspond to prognostic predictions and interventions in the clinical-medical models described in Section 1. Of the six determinants, cost-benefit and risk cognition are considered to be broad-sense utilities. Moreover, social norms and responsibility are considered to be broad-sense norms. These correspond to the issues of balance between utilities and norms in decision-making described at the beginning of the previous section. Finally, intervention in norms is linked to interventions in mutual expectations and approval, as described in Section 3, and it affects individual behaviour through inter-individual interaction as shown in Figure 8.

## 4.3 | Slow loop: Group intention diagnostics

In group intention diagnostics, it is necessary to know the pluralistic value index concerning the group in advance as the first step. This corresponds to the prognostic predictions of the clinical medicine model described in Section 1 and can be regarded as an assessment of evidence for the group consensus building discussed in the next section. However, it is difficult to determine the complex relationships between indicators involving various stakeholders using manual calculations. Therefore, we decided to utilise numerical simulation technology to calculate a pluralistic value index [36].

As shown in Figure 9, we took up local production and consumption of energy in communities as an example [37] and conducted a multi-agent simulation consisting of residents, local business operators, electric power companies etc. as stakeholders, and solar and hydraulic power plants, transmission and distribution grids, and power storage facilities etc. as energy facilities. Thus, we obtained approximately 20,000 calculation results, that is, choices, by changing the parameters. Among the numerous indices, the regional economic circulation rate (regional activation) was chosen for social value, natural energy utilisation rate for environmental value, and energy cost (household expenditure) for economic value. The calculated results were normalised and plotted in a ternary graph [38].

In the second step, group intention is obtained by aggregating the subjective preferences of individuals. Traditionally, Condorcet's Paradox and Arrow's Impossibility Theorem are known for the aggregation of preferences for multiple choices. However, it is not possible in the first place for an individual to determine the order of preferences for all choices shown in Figure 9. Therefore, it is necessary to consider a more practical method than raising the issues of Condorcet's Paradox or Arrow's Impossibility. Thus, we devised the paired comparison learning method (PCLM) to correspond individual preferences to network graphs of ternary values by repeating paired comparison questions, allowing individuals to judge preferences and machine-learning their answers, as illustrated in Figure 10 [39].

In the third step, group intention is diagnosed by mapping the preferences of the population obtained by PCLM in the second step against the large number of choices obtained in the first step. Specifically, based on the preference ratio between the two in the ternary graph (social, environmental, and economic value), the centroids derived from the geometric calculations shown in Figure 11a are plotted on the ternary graph as individual preferences. It is also possible to group a collection of individual preferences, as shown in Figure 11b, to explicitly indicate, for example, the majority (red dots group) and minority (blue dots group) intentions. In the future, estimating the comparison preferences for ternary values in addition to personalities in the diagnostics of individual behaviour in Section 4.1 may allow for the automatic diagnosis of group intention without questioning individuals.

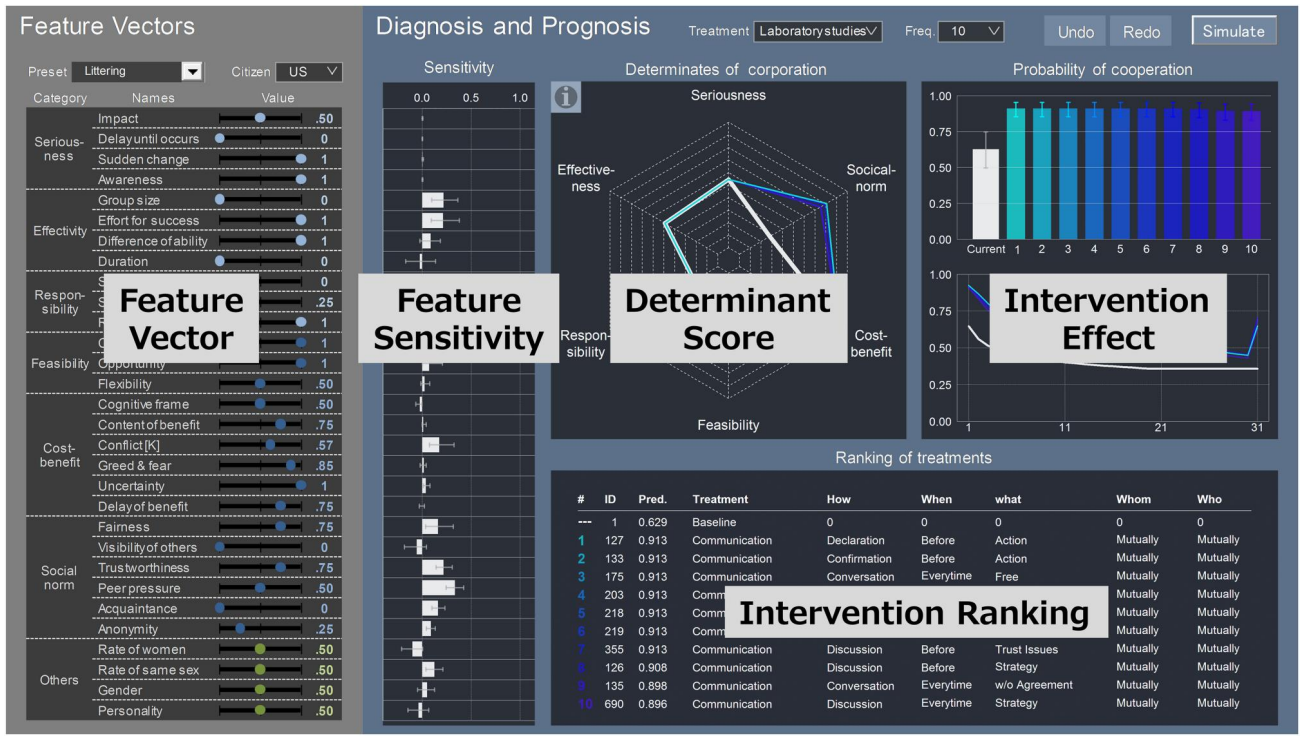

**FIGURE 7** Individual behavioural intervention dashboard

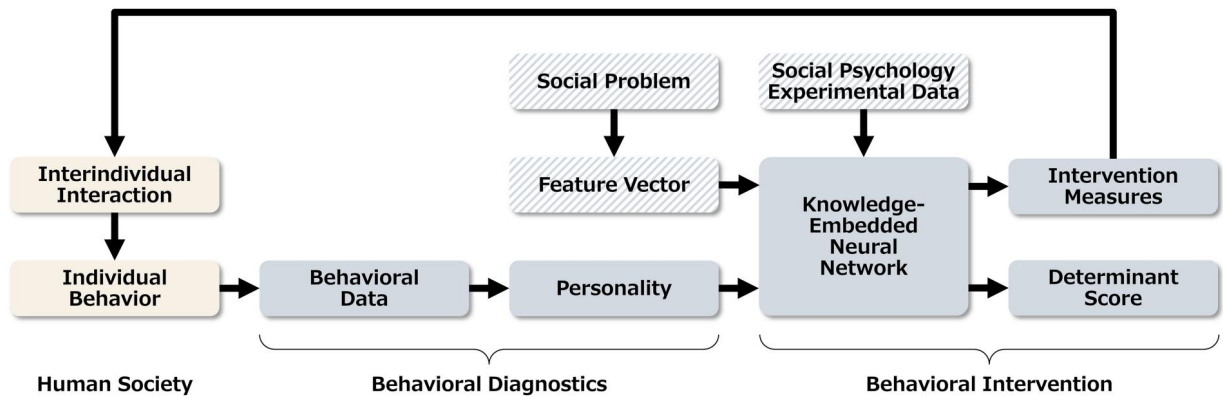

**FIGURE 8** Fast loop configured by individual behavioural diagnostics and individual behavioural intervention

## 4.4 | Slow loop: Group consensus building

Democracy (majority rule) emerged from the military requirements of ancient Greece, but the formation of an egalitarian consensus has been valued in 'mixed-life societies' consisting of multiple ethnic groups such as ancient Ionia and Zomia in Southeast Asia [40]. Game theory also proves the existence of a process leading to consensus for all, and social choice theory presents Rawls' difference principle (the greatest benefit of the least advantaged) for Bentham's principle of utility (the greatest happiness of the greatest number).

Considering these, the group consensus building of the slow loop aims not to mandate consensus by majority rule but to agree through the procedural fairness process (to elicit compromises that are not enough to reject consensus by anyone). Therefore, we decided to present a compromisable relationship that could mitigate conflict, along with a consensus reference point and a conflicted relationship between groups, as shown schematically in Figure 12a, based on the assessment of pluralistic value and the group intention diagnostics in the previous section [41].

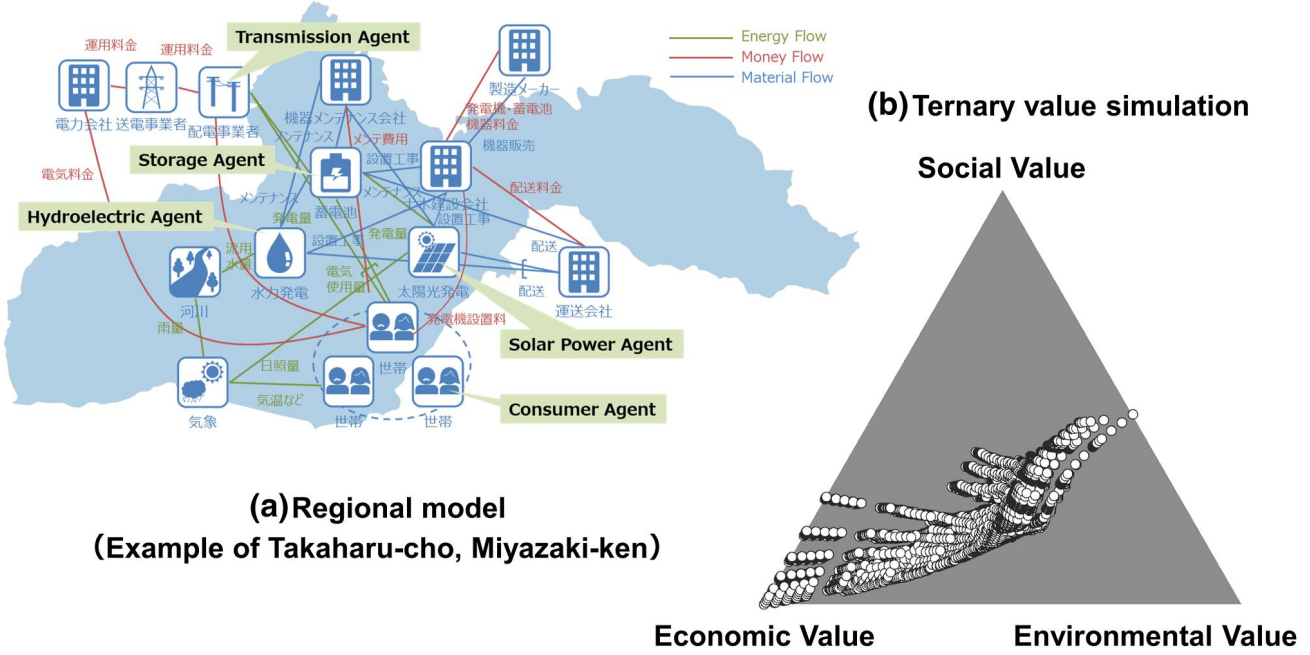

**FIGURE 9** Ternary value assessment

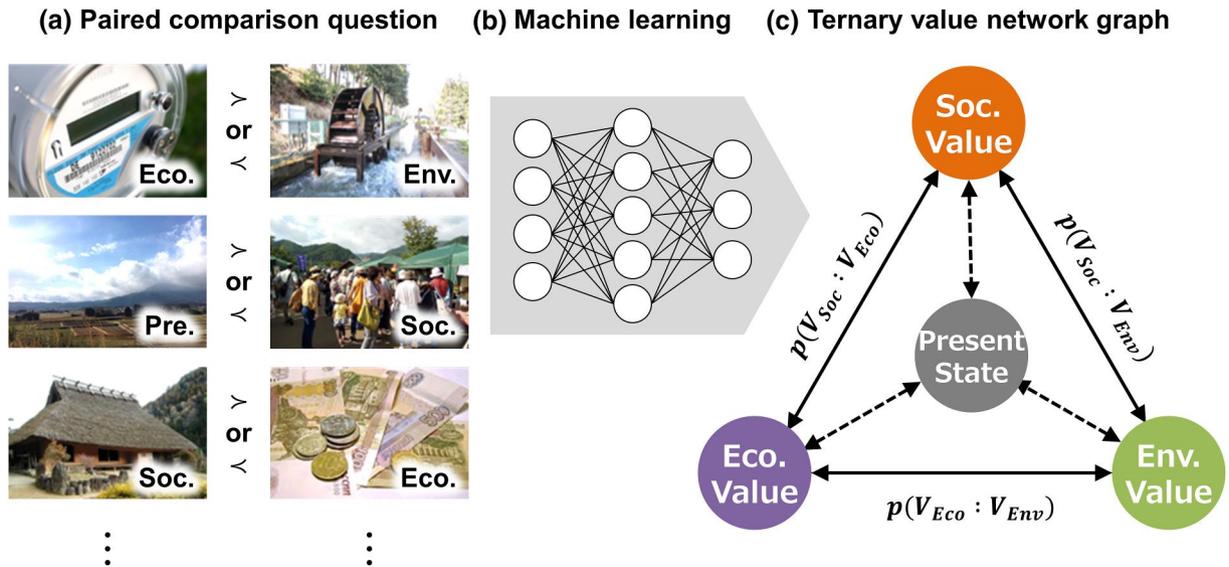

**FIGURE 10** Paired comparison learning method (PCLM)

Specifically, we consider the linear pathway connecting the centroid of preferences of all groups (consensus reference point) and the centroids of preferences of each group (preference aggregation points) as conflict relationships. Additionally, we consider the bypassing pathway formed by the auxiliary lines parallel to the edges of the ternary graph drawn from the preference aggregation points as compromisable relationships. In a compromisable relationship, one value is constant, and only weak transitivity is in a two-value comparison, which is expected to make it easier for participants to discuss. Thus, it is likely that conflicting groups will reach a consensus by changing each other's preferences in a relationship. This corresponds to progressively narrowing down the range (grey area) within the compromisable relationship shown in Figure 12a.

In general, the consensus-building process consists of five steps: convening (gathering participants and initiating assessments), clarifying roles and responsibilities (sharing of stakeholders and facilitators), facilitating group problem solving, achieving consensus (confirming consensus for all), and implementing commitment (implementing consensus content and re-convening according to context change) [42]. The presentation of the conflict and the compromisable relationships described in this section will support the third step.

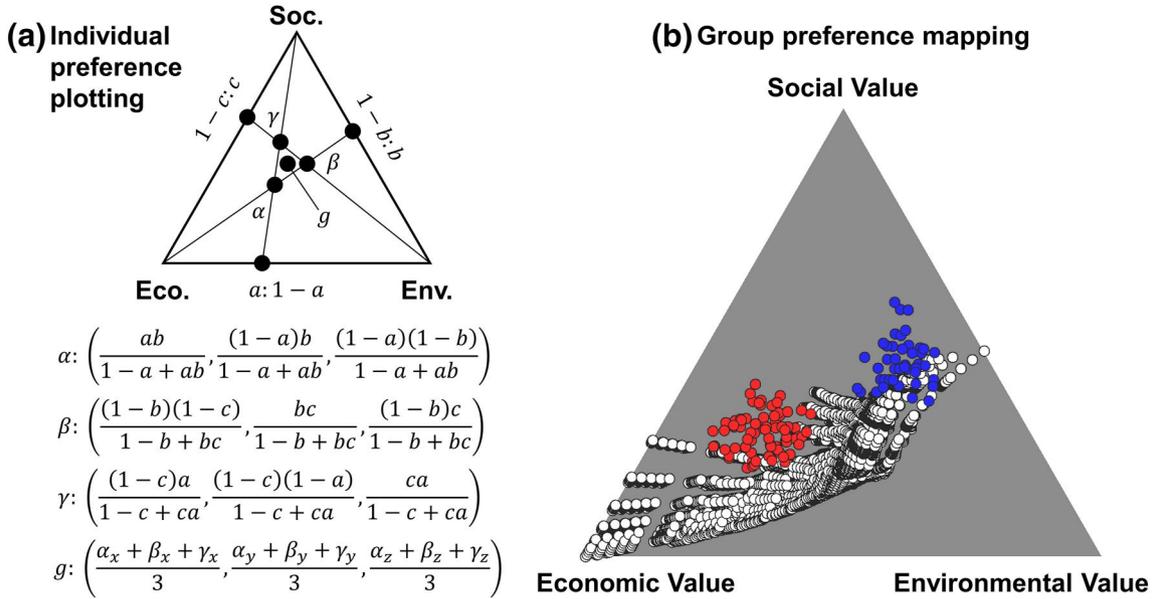

**FIGURE 11** Group preference mapping

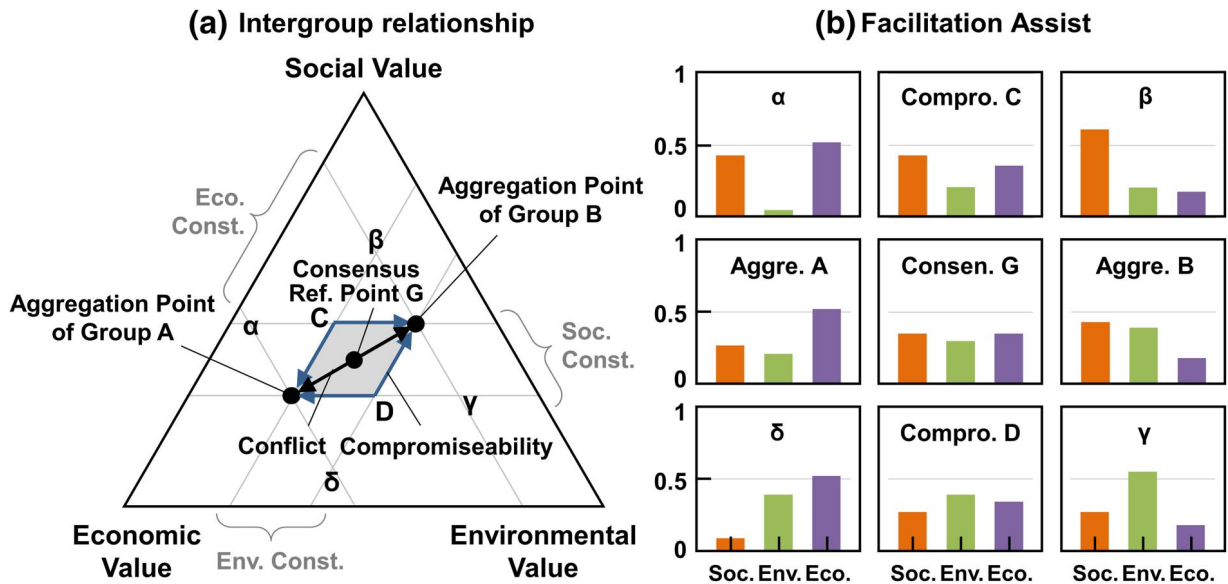

**FIGURE 12** Presentation of intra-group relationship

Facilitators will demonstrate the position of the group (preference aggregation points), conflict relationship, and compromisable relationships, prompting each other's compromise while explaining specific values and the details of the choices, as shown in Figure 12b.

The final consensus draft does not necessarily need to occur at the consensus reference point or the midpoint of the compromisable pathway. For example, when considering disadvantaged minorities, a choice leaning closer towards the minority preference rather than a reference point or a midpoint would guide the proposed consensus as a social choice. Similar to our proposed aggregation method for weighting minority positionality [43], we could support social choice by weighting the majority-to-minority ratio (majority/minority) and the pluralistic value dimensionality ratio (total dimension/minority-respected dimension).

In the slow loop, we believe that consensus building can be more effectively supported by connecting the group intention diagnostics information presented in the previous section to the group consensus-building facilitation information presented in this section, as shown in Figure 13. Ternary value assessment, intention diagnostics, and consensus building support correspond to prognostics, diagnostics, and intervention in the clinical medicine model described in Section 1, respectively. In addition, as discussed in Section 1, consensus building in a social system influences individual behaviour through social institutions, resulting in social order.

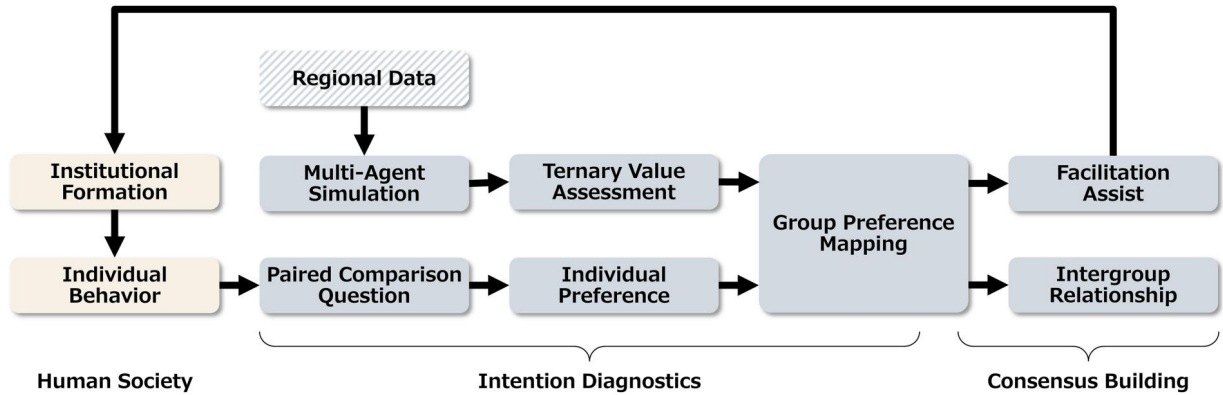

**FIGURE 13** Slow loop configured by group intention diagnostics and group consensus building

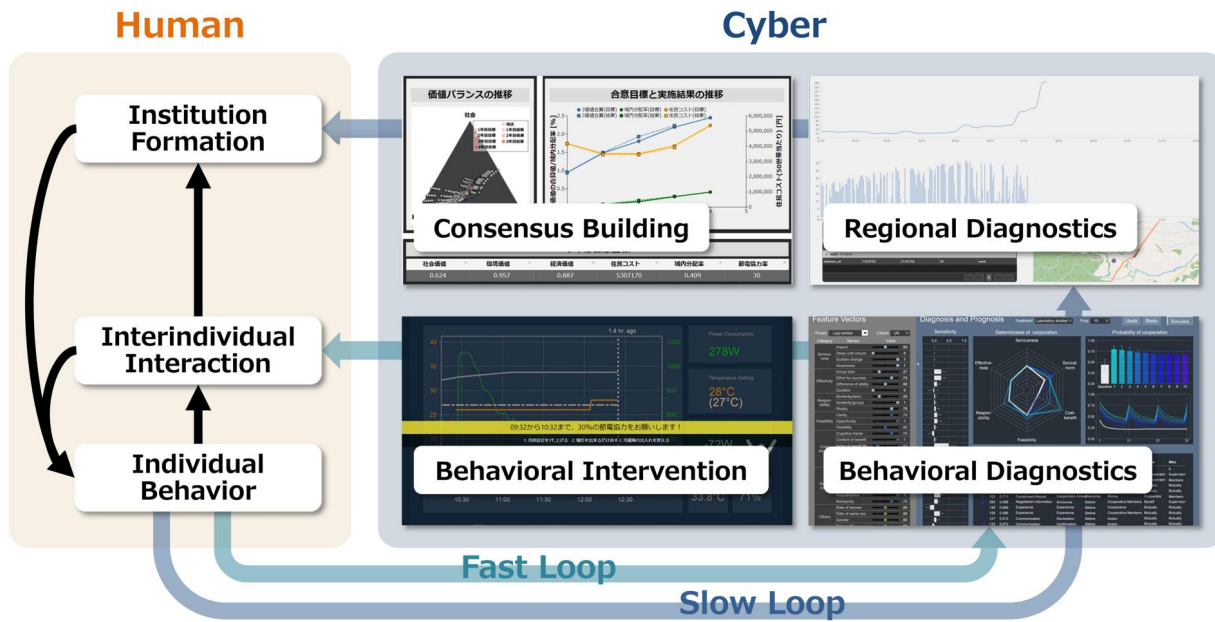

**FIGURE 14** Cooperative operation of Social Co-OS

The entire Social Co-OS can be constructed by connecting the four basic functions described in this section. An example of a demonstration of the cooperation between fast and slow loops is shown in Figure 14 (for details and movies of the demos, please refer to public websites [44]). Here, the production and consumption of energy in a local community are exemplified. In the individual behavioural diagnostics of the fast loop, intervention measures for energy conservation are presented, and in the individual behavioural intervention, an intervention that promotes energy conservation in proportion to power consumption is performed. In the regional diagnostics of the slow loop, the transition of power generation by solar or hydraulic power is presented; in the group consensus building, the agreement on the ternary value is reviewed according to the power generation situation.

## 5 | DISCUSSION

In the fast loop of Social Co-OS, more effective behavioural interventions could be provided with personal attribute information (e.g. age, gender, and ethnicity) and personality traits based on psychological tests and behavioural observational data. Similarly, the slow loop would facilitate consensus building more efficiently if, for example, a person's social status (e.g. occupation, job, and annual income) and a human relationship network based on positional information data were obtained. Specifically, there is a trade-off between personal information protection and the effects of behavioural interventions and consensus building. Although consent to handling personal information is a prerequisite for the participation phase of Social Co-OS, consideration must be given to security and privacy protection, portability and deletion rights, biases and equity of collected data, transparency and explanation, as well as discussions on AI ethics. However, the Social Co-OS only supports behavioural interventions and consensus building, eventually depending on decision-making by individuals and groups. Additionally, ethical issues, including personal information, seem to result in empathy and trust in the Social Co-OS community [41].

Similar to behavioural economic nudges, there are concerns about individual behavioural interventions as well as paternalistic interference. However, interventions may be acceptable if participation in the Social Co-OS is based on free will and informed consent. Additionally, from the viewpoint of 'naturalness' and 'de-ethics,' which is an East Asian human perspective [45], the social norm in the intervention will be changed from a uniform 'should' rule (control of variability) to a 'could' rule (mutual approval of variability). It includes diversity to aim for a 'mixed-life society' characterised by mutual entrustment. Additionally, selfish-free riders may not gain mutual approval, so the Social Co-OS does not fall into disorder. This may have replaced the graduated sanctions in Ostrom's principles for managing commons [46] with softer rules.

In terms of group consensus formation, the distinction between minorities and egoists remains a challenge in achieving group consensus. Regarding this task, if consent has been obtained at the participation stage for shared goals, such as SDGs, selfish subjective preferences could be visualised in group intention diagnostics. Furthermore, the egoists and others around them could be made aware of the deviation from the shared goals, thus prompting the egoists to make a compromise. The liberal trilemma exhibited by Bowles cannot simultaneously meet a tripartite consisting of Pareto efficiency, voluntary participation, and preference neutrality [47]. Promoting compromise to shared goals leads to suppressing egoist preference neutrality and emphasising Pareto efficiency and voluntary participation in Social Co-OS.

## 6 | CONCLUSION

Towards solving social challenges and realising a 'mixed-life society,' this study reviewed current visions of Society 5.0, proposed the concept of CHSS in which the cyber and human societies operate in mutual commitment, and presented the architecture and basic functions of Social Co-OS. These were achieved by an interdisciplinary fusion of information science with the humanities and social sciences concerned with information systems and social systems.

As a future development, the function of Social Co-OS should be expanded. For example, the Social Co-OS incorporation of emotion analysis technologies that utilise multimodal information, such as facial expressions, voice, and gestures, in individual behavioural diagnostics and multimodal interaction technologies via graphical interfaces and robots in individual behavioural interventions could allow for more effective and affiliative interventions.

Additionally, we will continue to combine new technology with existing consensus-building support tools for the intention diagnostics and consensus-building groups. For example, the Decidim democratic platform for citizen participation [48] and the Loomio online organisational decision-making tool [49] can perform various functions, such as proposals, assessments, meetings, discussions, bulletin boards, voting, and choices. Combining these functions with the pluralistic value simulation technology, group preference mapping, and representation of intra-group relationships presented in this study allows for more efficient consensus building.

It should be noted that CHSS and Social Co-OS, which aim for mutual entrustment of 'mixed-life societies,' are committed to improving the wellbeing of humans. In positive psychology, role-playing and prosocial games enhance psychological well-being through empathy and compassion [50]. In studies of alternate reality games, massively multiplayer games enhance social connectedness and cooperative behaviours [51]. Thus, it would be effective to incorporate gamification in the context of mutually approved behavioural interventions and consensus building in Social Co-OS.

In the future, Social Co-OS's social practices will cover mutual aid societies, such as local governments and communities. The European Union is driving data democratisation, such as the General Data Protection Regulation, and aims to advance digital democracy [52] and platform cooperativism [53]. The former cites freedom and equality in digital technology, while the latter refers to joint ownership and the democratic governance of information platforms. The CHSS and Social Co-OS presented in this study will contribute to these movements.

## AUTHOR CONTRIBUTIONS

**Takeshi Kato**: Conceptualization, Investigation, Methodology, Project administration, Supervision, Visualization, Writing – original draft, Writing – review & editing. **Yasuyuki Kudo**: Data curation, Formal analysis, Investigation, Methodology, Software, Validation, Visualization, Writing – review & editing. **Junichi Miyakoshi**: Data curation, Formal analysi, Investigation, Methodology, Software, Validation, Visualization, Writing – review & editing. **Misa Owa**: Data curation, Formal analysis.

## ACKNOWLEDGEMENTS

This study was conducted in a collaborative research department established by Kyoto University and Hitachi Ltd. We would like to thank Professor Yasuo Deguchi of Kyoto University, Associate Professor Jun Otsuka of Kyoto University, Professor Hayato Saigo of Nagahama Institute of Bio-Science and Technology for providing suggestions from the viewpoint of philosophy, Professor Kaori Karasawa of Tokyo University, Professor Hiroyuki Yamaguchi of Kyushu University for providing suggestions from the viewpoint of social psychology, and Professor Yoshinori Hiroi of Kyoto University for advising on social concepts and social experiments in communities. In addition, we would like to thank Editage (www.editage.com) for English language editing.

## CONFLICT OF INTEREST

The author declares that there is no conflict of interest that could be perceived as prejudicing the impartiality of the research reported.

## DATA AVAILABILITY STATEMENT

The data that support the findings of this study are available from the corresponding author upon reasonable request.

## ORCID

*Takeshi Kato* https://orcid.org/0000-0002-6744-8606
*Yasuyuki Kudo* https://orcid.org/0000-0001-6476-8148
*Junichi Miyakoshi* https://orcid.org/0000-0003-2989-7619
*Misa Owa* https://orcid.org/0000-0002-2570-2354
*Yasuhiro Asa* https://orcid.org/0000-0002-0349-6537
*Takashi Numata* https://orcid.org/0000-0001-6981-1766
*Ryuji Mine* https://orcid.org/0000-0002-0130-6752
*Hiroyuki Mizuno* https://orcid.org/0000-0002-1213-9021

## SUPPORTING INFORMATION

Additional supporting information can be found online in the Supporting Information section at the end of this article.